\documentclass[pre,aps,twocolumn,amsmath,amssymb]{revtex4}

\pdfoutput=1
\usepackage{hyperref}
\usepackage{graphicx}

\def\eqq#1{Eq.~(\ref{#1})}
\def\eq#1{(\ref{#1})}

\def\f#1{Fig.~\ref{#1}}

\def\c#1{~\cite{#1}}

\def\en{\epsilon_{\rm d}}
\def\enn{\epsilon_{\rm u}}

\def\ben{\overline{\epsilon_{\rm d}}}
\def\benn{\overline{\epsilon_{\rm u}}}

\def\den{{\Delta_{\rm d}}}
\def\denn{{\Delta_{\rm u}}}

\def\e{{\rm e}}

\def\beq{\begin{equation}}
\def\eeq{\end{equation}}
\def\bea{\begin{eqnarray}}
\def\eea{\end{eqnarray}}

\def\kt{k_{\rm B}T}

\begin{document}

\title{Growth of equilibrium structures built from a large \\number of distinct component types}

\author{Lester O. Hedges, Ranjan V. Mannige, Stephen Whitelam\footnote{\texttt{swhitelam@lbl.gov}}}
\affiliation{Molecular Foundry, Lawrence Berkeley National Laboratory, 1 Cyclotron Road, Berkeley, CA 94720, USA}
\begin{abstract}
We use simple analytic arguments and lattice-based computer simulations to study the growth of structures made from a large number of distinct component types. Components possess `designed' interactions, chosen to stabilize an equilibrium target structure in which each component type has a defined spatial position, and `undesigned' interactions that allow components to bind in a compositionally-disordered way. We find that high-fidelity growth of the equilibrium target structure can happen in the presence of substantial attractive undesigned interactions, as long as the energy scale of the set of designed interactions is chosen appropriately.  This observation may help explain why equilibrium DNA `brick' structures self-assemble even if undesigned interactions are not suppressed [Ke et al. Science 338, 1177 (2012)]. We also find that high-fidelity growth of the target structure is most probable when designed interactions are drawn from a distribution that is as narrow as possible. We use this result to suggest how to choose complementary DNA sequences in order to maximize the fidelity of multicomponent self-assembly mediated by DNA. We also comment on the prospect of growing macroscopic structures in this manner.
\end{abstract}
\maketitle

\section{Introduction}

\begin{figure*}[t]
\centering
\includegraphics[width=0.8\linewidth]{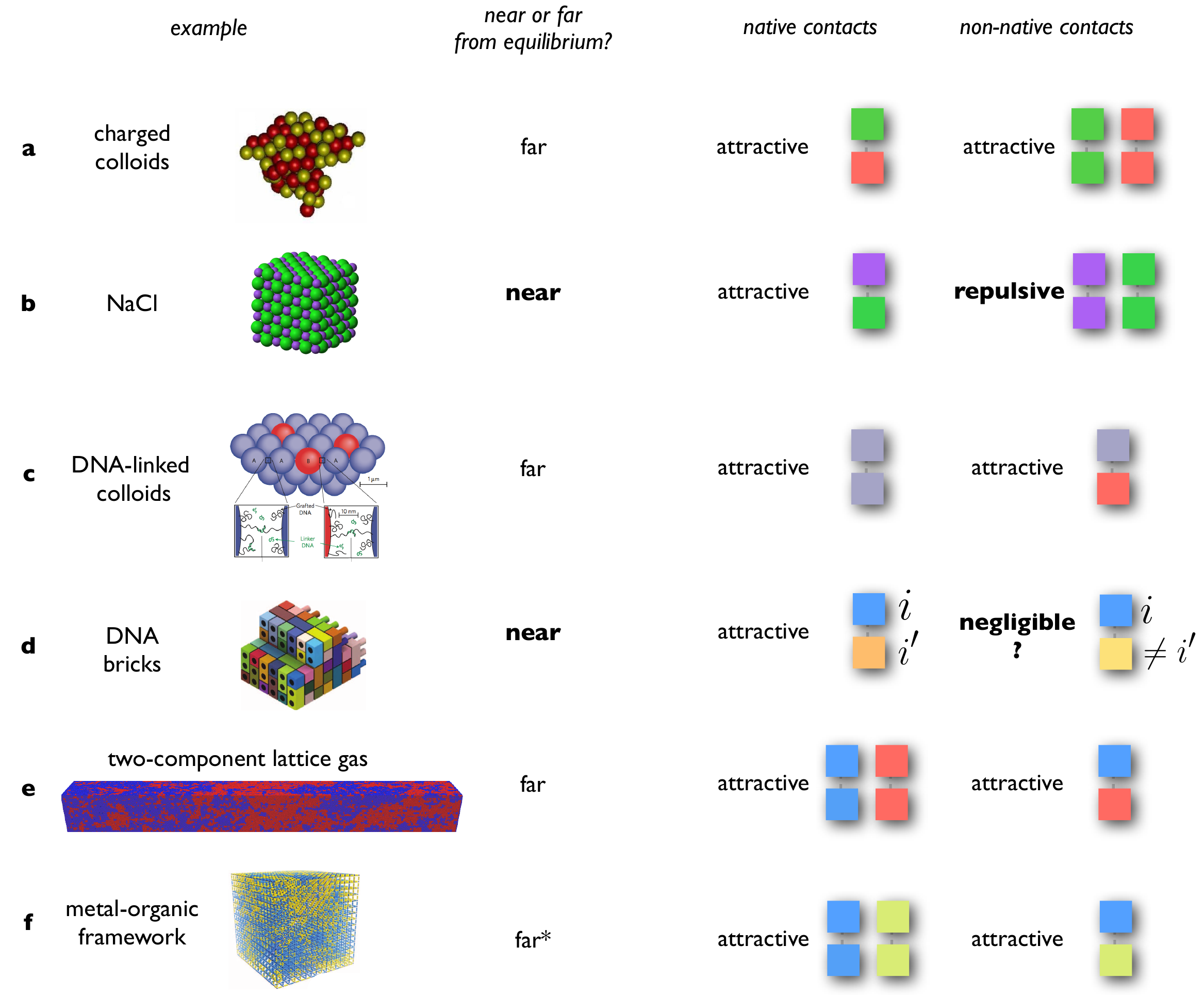}
\caption{\label{fig1} Examples of self-assembled multicomponent structures `near' and `far from' equilibrium. For the examples shown we distinguish `native' and `non-native' interactions, which are respectively abundant and rare in the equilibrium structure.  However, the component type arrangement that results from self-assembly may contain many non-native contacts that are kinetically trapped within a solid structure. Examples (a), (c), (e) and (f) show nonequilibrium structures that result from such kinetic trapping: in these structures, component types possess a spatial arrangement not compatible with thermodynamic equilibrium. The examples shown are charged colloids on the computer\c{sanz2007evidence}, DNA-linked colloids in experiment\c{kim2008probing}, a two-component lattice gas on the computer\c{whitelam2014self}, and a computer representation of an experimental metal-organic framework\c{kong2013mapping} (the asterisk indicates that the `far from equilibrium' designation is our interpretation of the results described in that paper). By contrast, examples (b) and (d), sodium chloride (image from\c{nacl}; see also Ref.\c{valeriani2005rate}) and DNA `bricks'\c{ke2012three}, are examples of multicomponent self-assembly that do result in the equilibrium structure, or something very close to it. For the case of NaCl it is clear that non-native contacts are suppressed because they are repulsive, so explaining why the equilibrium structure forms without difficulty during self-assembly. The DNA brick case appears to be different, in that its non-native interactions are not guaranteed to be negligible. Here we show, within a simple computer model, that the growth of equilibrium structures of a large number of component types can occur even in the face of substantial attractive non-native (henceforth called `undesigned') interactions between component types, as long as native (henceforth called `designed') interactions are well-enough separated from them in energy scale. For image permissions, see end of paper.}
\end{figure*}

The self-assembly of multicomponent solid structures can be affected by kinetic traps that emerge even under mild nonequilibrium conditions: the slow swapping of component types within solid structures can prevent particle types from achieving their equilibrium spatial arrangement as the structure they comprise nucleates and grows\c{kremer1978multi,stauffer1976kinetic,PhysRevB.27.7372,schmelzer2004nucleation,schmelzer2000reconciling,scarlett2010computational,kim2008probing,scarlett2011mechanistic,sanz2007evidence,peters2009competing,whitelam2012self}. Some examples of this phenomenon include the crystallization of charged colloids on the computer\c{sanz2007evidence,peters2009competing}, and of DNA-linked colloids in experiment and simulation\c{scarlett2010computational,kim2008probing,scarlett2011mechanistic}. In these examples, self-assembly results in ordered crystal structures harboring nonequilibrium arrangements of component types: see \f{fig1}.

Of course, many examples of multicomponent self-assembly do not encounter such kinetic trapping: multiple component types can also self-assemble into equilibrium structures. In order to ensure self-assembly of the equilibrium structure it would seem to be sufficient to choose the energy scales of component-type interactions so that `non-native' bonds, i.e. bonds seen with low likelihood in the equilibrium structure, also form with low likelihood during self-assembly. Consider solid sodium chloride: the non-native like-charge bond is repulsive, and so its formation during crystallization is much less likely than is the formation of a `native' unlike-charge bond\c{valeriani2005rate}. But is it strictly necessary, in order to achieve assembly of the equilibrium structure, to make non-native interactions repulsive? The remarkable self-assembly of equilibrium DNA brick structures\c{ke2012three} consisting of ordered arrays of distinct component types would seem to suggest not, because assembly of the equilibrium `target' structure happens in the presence of potential `undesigned' attractive interactions between component types (i.e. attractions between component types that are {\em not} neighbors in the target structure). A recent lattice-based simulation study of the nucleation and growth of multicomponent `patchy colloids'\c{reinhardt2014numerical} shows that multicomponent self-assembly of this nature can be achieved by mimicking only the general sense of DNA complementarity, without accounting for fine details of the experimental system. In addition, recent off-lattice simulations\c{halverson2013dna} show that particles possessing similar component-type complementarity can self-assemble into equilibrium structures of arbitrary shapes.
\begin{figure*}[]
\centering
\includegraphics[width=0.85\linewidth]{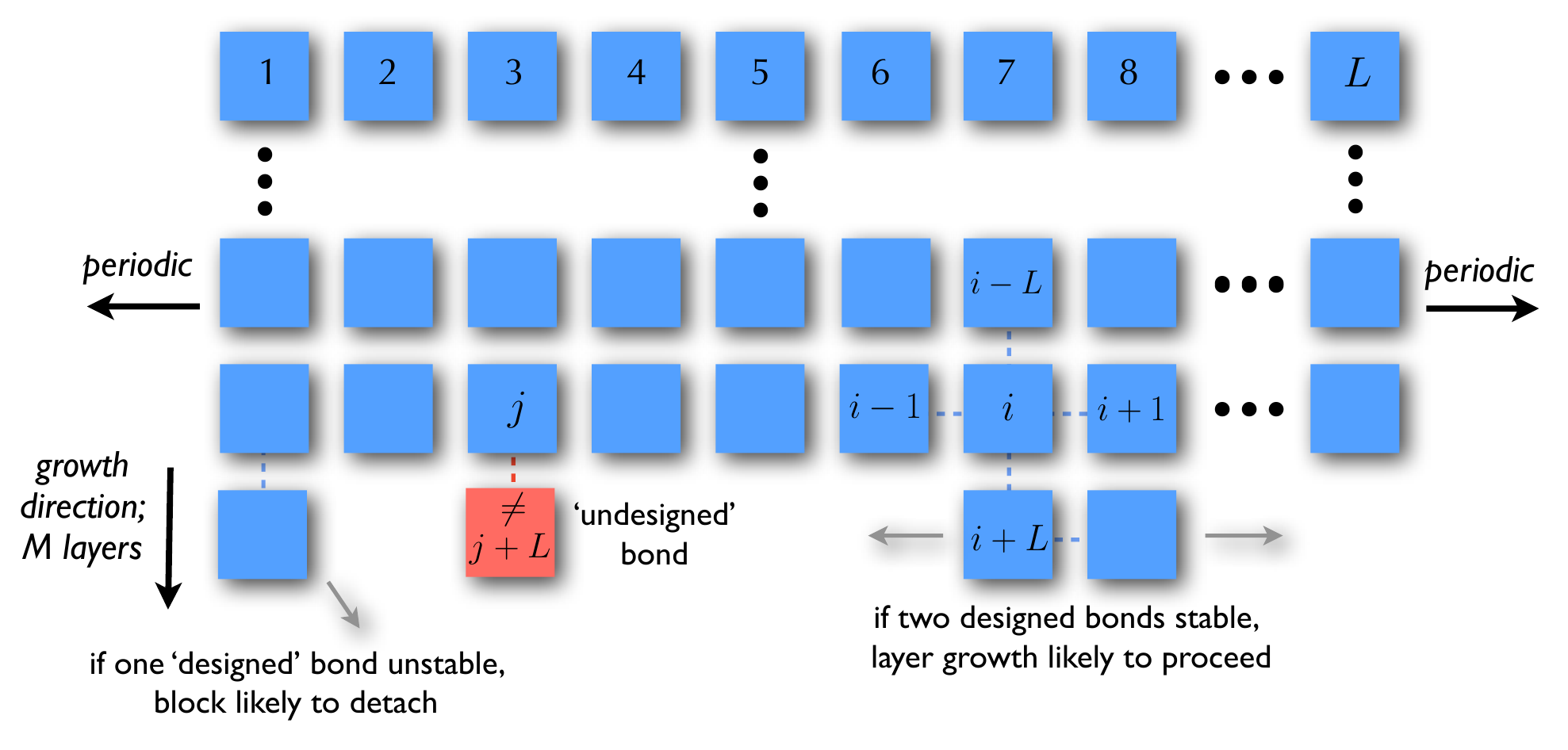}
\caption{\label{fig2} Schematic of the growth process considered in this paper. Blocks of $Q$ distinct types bind in a `designed' way according to the code shown in the vicinity of the block of type $i$, and in an `undesigned' way otherwise (see red block). The template of width $L$ shown on the top row promotes the growth of a structure the downward direction. We define the `ideal' structure as an assembled structure in which block types run from 1 to $L$ on the first row, from $L+1$ to $2L$ on the second row, etc., and from $(M-1)L+1$ to $ML$ on the bottom row. For a range of parameters the thermodynamically stable structure can be arranged to be very close to the ideal structure; this stable structure is then the `target' structure for self-assembly. Simple conditions applied to designed and undesigned interaction strengths allow the self-assembly of this equilibrium target structure with high fidelity (see text).}
\end{figure*}

Taken as a whole, these studies\c{ke2012three,halverson2013dna,reinhardt2014numerical} suggest that the self-assembly of equilibrium multicomponent structures of arbitrary complexity might be possible quite generally, provided than one can arrange to have particular component-type interactions. But just how precisely must component-type interactions be controlled in order to have the equilibrium structure self-assemble? Here we address this question within a lattice-based computer model of $Q \sim 10^3$ distinct component types. Components interact and self-assemble via `designed' and `undesigned' interactions (Section \ref{model}). The former stabilize a defined target structure in which each component type has a prescribed spatial position, while the latter allow component types to associate in a compositionally-disordered way.  By varying these interactions we determine where in parameter space one can grow an equilibrium multicomponent structure of arbitrary component type arrangement. We use a `template' to seed growth without waiting for spontaneous nucleation, 
and so our treatment of this self-assembly problem is only partial: we do not attempt to determine how to nucleate a multicomponent structure (see e.g.\c{reinhardt2014numerical}). Nor do we determine how to grow multicomponent structures of defined shapes (see e.g.\c{licata2006errorproof,halverson2013dna}). Instead, we determine how best to grow simple rectangular shapes composed of precisely-arranged component types, motivated by the observation that if one wishes to self-assemble large equilibrium structures then one must arrange for the growth phase of self-assembly to happen `near' to equilibrium.

In Section~\ref{results1} we show that, in the absence of undesigned interactions, high-fidelity growth of the equilibrium target structure is most probable when designed interactions are drawn from a distribution that is as narrow as possible. Otherwise, different pieces of the target structure have a tendency to form in different regions of parameter space, making assembly of the complete target structure less probable. In Section~\ref{results2} we show that growth can also happen with high fidelity in the presence of substantial attractive undesigned interactions, provided the energy scales of designed and undesigned interactions are sufficiently separated. In Section~\ref{discussion} we discuss these findings in the context of DNA-mediated interactions, and suggest how to select complementary DNA sequences in order to maximize the success rate of assembly mediated by DNA. Example sequences are given as \href{http://nanotheory.lbl.gov/people/dna_sequences/}{supplementary files}. We conclude in Section~\ref{conclusions} by commenting on the prospect of growing equilibrium multicomponent structures as the number of component types becomes very large.

\section{Model and Simulation Methods}
\label{model}

\noindent {\bf Model.} We consider a generalization, sketched in \f{fig2}, of the lattice model growth procedure used in Ref.\c{whitelam2014self}. The lattice is a 2D square one of $M \times L$ sites. Sites may be unoccupied, or occupied by a block of type $i$, where $i=L+1,L+1,\dots,2,\dots,Q$, and $Q=LM$ is the total number of block types (block types $i=1,2,\dots,L$ are reserved for a top-row `template' whose purpose is to promote growth; see below). Block interaction energies are nearest-neighbor ones, of two types: `designed' and `undesigned'. Designed interactions occur only between one face of one block type and one face of a second, particular block type, according to the following rules (see \f{fig2}): block type $i$ can make 4 designed nearest-neighbor bonds, one with block type $i+1$ to the immediate right, one with block type $i-1$ to the immediate left, one with block type $i-L$ immediately above, and one with block type $i+L$ immediately below. Thus if a block of type $i+L$ sits immediately to the {\em right} of a block of type $i$, no designed bond would be made. In this way, one can arrange on the lattice an $M \times L$ grid of blocks, all making 4 designed bonds, modulo boundary conditions:  we imposed periodic boundary conditions in the horizontal direction of the lattice, and closed boundary conditions in the vertical direction of the lattice.  Each pairwise designed bond brings with it an energy reward $-\en \kt$, where $\en$ is a random number chosen from a Gaussian distribution with mean $\ben$ and variance $\den$. The random numbers required to define all possible designed interactions are chosen at the start of each simulation.
\begin{figure*}[]
\centering
\includegraphics[width=\linewidth]{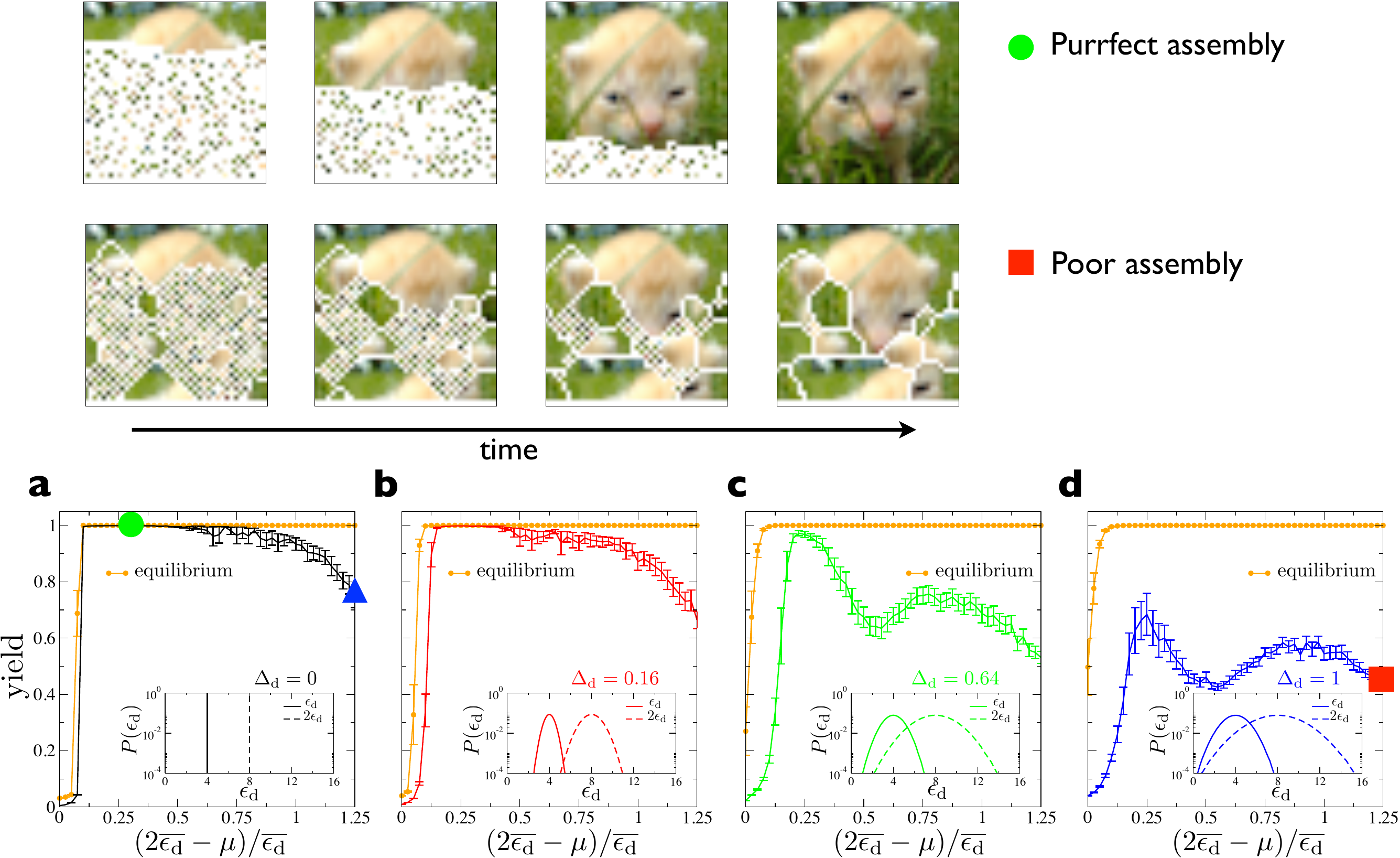}
\caption{\label{fig3} Results of growth simulations in the presence of designed interactions only. {\em Top}: Symbols relate the displayed time series to the parameters at which they were calculated (see panels (a) and (d)). We associate block types in the target structure with pixels, arranged to form an arbitrarily-chosen image\c{kitty}; this image provides a guide to the eye as to the fidelity of assembly. {\em Bottom}: We plot equilibrium yield (orange) and dynamic yield taken at $10^7$ timesteps for growth simulations done at different values of chemical potential $\mu$ (horizontal axes). Each curve has been averaged over 200 independent trajectories and error bars indicate the standard deviation at each data point. The combination $(2\ben-\mu)/\ben$ is chosen so that the window of good assembly predicted for the case in which all designed interactions are equal in strength, defined by \eqq{window1}, corresponds to a range from 0 (weak driving, slow growth) to 1 (strong driving, fast growth) on the horizontal axis. In panel (a) we see that the prediction of \eqq{window1} is reasonably accurate: when designed interactions are drawn from a distribution with zero variance (see inset), high-fidelity growth of the 
target structure happens within most of the expected window. The window of high-fidelity growth narrows and eventually disappears as the designed interaction variance widens, as shown in panels (b) to (d). Thus, even if undesigned interactions are completely suppressed, designed interactions must be drawn from a narrow-enough distribution in order for the equilibrium structure to grow directly. Otherwise, far-from-equilibrium growth of fragments of the desired structure happens.}
\end{figure*}

Blocks also make undesigned bonds with nearest-neighbor blocks, again in an orientation-dependent fashion, with any block type {\em except} the one that would give rise to a designed bond. Thus if a block of type $i$ sits to the right of a block of type $j$ (where $j$ is any number {\em except} $i-1$) then a undesigned bond is made; if a block of type $i$ sits just below a block of the same type $j$ (as long as $j \neq i-L$) then a different undesigned bond is made. Each pairwise undesigned bond brings with it an energy reward $-\enn \kt$, where $\enn$ is a random number chosen from a Gaussian distribution with mean $\benn$ and variance $\denn$. The random numbers required to define all possible undesigned interactions are chosen at the start of each simulation.

The top row of the lattice was a fixed `template' of block types arranged in a designed-binding fashion; these block types run, from left to right, $1,2,\dots,L$ (see \f{fig2}). The bottom row of the lattice has immediately below it a smooth boundary that does not interact energetically with any block type. We define the `ideal' structure as a fully-occupied lattice in which each block makes only designed bonds. Block types within this structure then run in row-by-row order from 1 at the top left to $Q$ at the bottom right. Blocks types that inhabit the final row of the ideal structure can make no designed interactions in the downward direction (regardless of where on the lattice those block types sit). We set $M=L=40$, giving $Q=1600$. 

We worked in the grand-canonical ensemble, modeling an infinite reservoir of all block types not in the template row. Blocks of all types received an energetic penalty, relative to an unoccupied site, of magnitude $\mu \, \kt$ (this is a chemical potential term, but note that positive $\mu$ {\em disfavors} particles relative to vacancies).\\

\noindent {\bf Equilibrium sampling.} We are interested in conditions for which the thermodynamically stable structure is very close to the ideal structure; this stable structure is then the `target' structure for self-assembly. We calculated the thermodynamically stable structure by building the ideal structure `by hand', and using the following Monte Carlo move to achieve equilibrium. We picked at random any lattice site not within the template, and proposed, with uniform probability, a change from that site's current state (any of $q \equiv Q-L$ block types or a vacancy) to any other state (a vacancy or any of the $q$ block types not contained within the template). We accepted this proposal with the usual Metropolis rate, $\min\left( 1, {\rm e}^{-\beta \Delta U}\right)$, where $\Delta U$ is the total energy change resulting from the proposed move (this energy includes, in general, contributions from bond energies and the chemical potential). Note that if one wished to maintain, in the absence of inter-block 
interactions, a fixed concentration of blocks (relative to vacancies) as $q$ is increased, then one must increase the chemical potential by an amount $\ln q$. We focus on this paper on the case of fixed $q$; in reading subsequent equations, it should be noted that if one were to imagine increasing $q$ and leaving $\mu$ unchanged, then one would also be imagining an increase of the notional solution concentration of blocks. 

To make the ideal structure thermodynamically stable with respect to the unoccupied lattice, one must make $\ben$ large enough relative to $\mu$: the (reduced) bulk free-energy density difference between the ideal structure and the vapor is roughly $\mu - 2 \ben $, because each particle in the bulk of the ideal structure has two full bonds (a half share of four bonds) and comes with an energetic penalty $\mu$. To ensure stability of the ideal structure with respect to a condensed phase of randomly-arranged block types we need to ensure that $\ben$ is large enough relative to $\benn$ and $\ln Q$: the bulk free-energy density difference between the ideal structure and an occupied lattice of essentially random block types -- ignoring template blocks -- is roughly $2\benn+\ln Q- 2 \ben$. \\

\noindent {\bf Dynamic protocol.} To model a dynamics that approximates growth of a structure from interacting building blocks that diffuse in solution, we did as follows. We chose at random a lattice site {\em not} in the top-row template. If unoccupied, we attempted to occupy the lattice site with a block of type $i=L+1, L+2,\dots,Q$, chosen uniformly from the ensemble of all possible block types {\em except} those contained within the template. We accepted this proposal with probability $\min\left( 1, q \, {\rm e}^{-\beta \Delta E- \mu} \right)$, where $\Delta E$ is the bond energy change upon the proposed insertion (recall that $\mu$ has already been scaled by $\beta \equiv 1/(\kt)$, and that $q \equiv Q-L$ is the total number of block types minus those found in the template). If the randomly-chosen lattice site was instead occupied, we proposed to remove the block found there, and accepted this proposal with probability $\min\left( 1, q^{-1} {\rm e}^{-\beta \Delta E+ \mu}\right)$, where $\Delta E$ is 
the bond energy change upon removing the particle from the simulation box. These rates satisfy detailed balance with respect to our chosen energy function. The factor of $q$ in the acceptance rates accounts for the following asymmetry. If an unoccupied site is chosen, then insertion of a particular block type is proposed with likelihood $1/q$; the reverse of that move, assuming that the same lattice site is chosen, is proposed instead with unit probability. To model the fact that relaxation within solid structures is slow, we imposed a kinetic constraint that prevents any change of state of a lattice site having exactly 4 occupied neighbors. This constraint prevents relaxation within the bulk of an assembly, except in the neighborhood of a vacancy. The constraint respects detailed balance, and so has no effect on the thermodynamics of the model.
\begin{figure}[t!]
\centering
\includegraphics[width=0.95\linewidth]{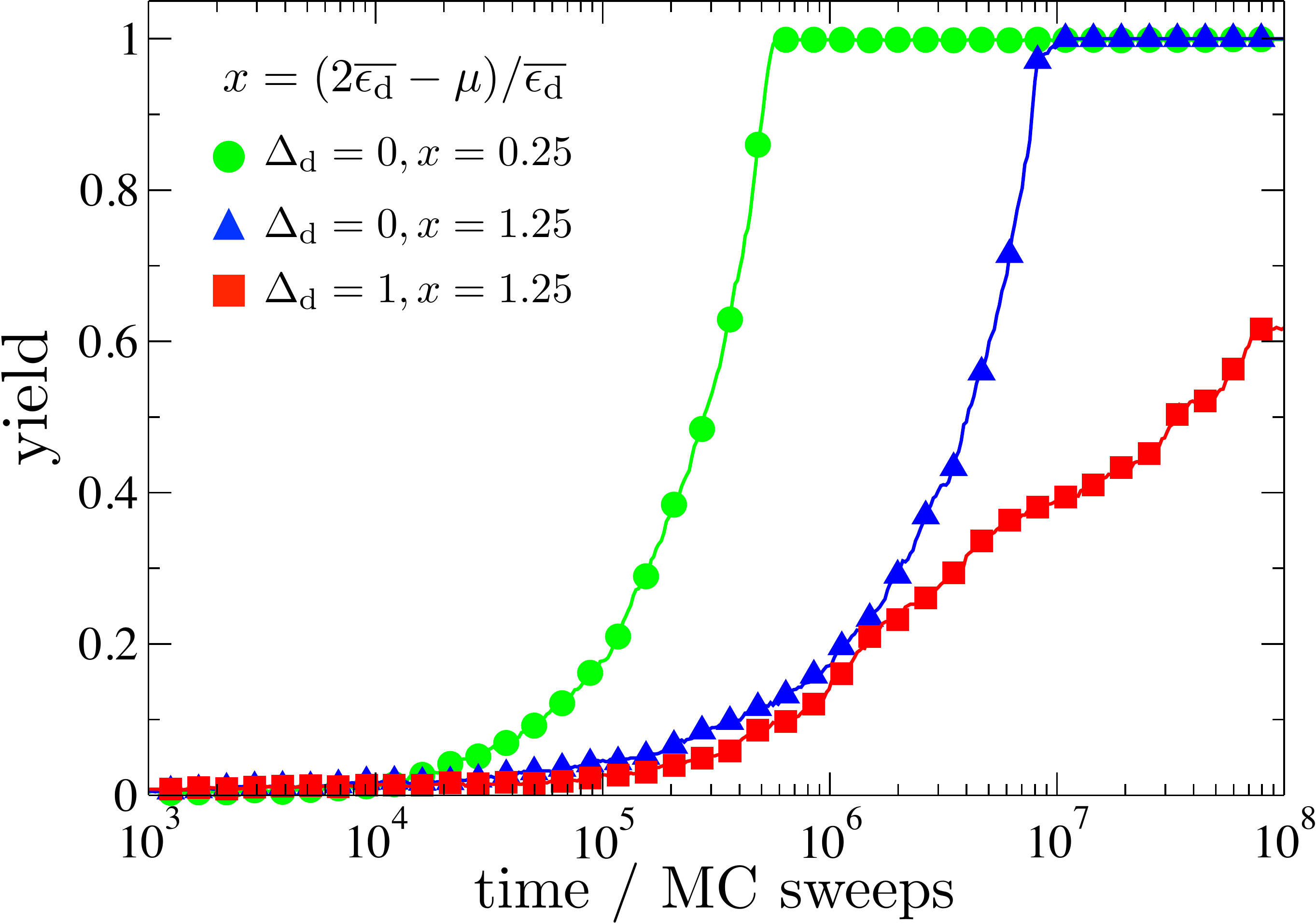}
\caption{\label{fig4} Yield-versus-time plots for examples of growth taken from three of the parameter sets used in \f{fig3}: see circle, triangle and square symbols on that figure.}
\end{figure}

Initially, the lattice was unoccupied apart from the top-row template. This template, similar to those used during the assembly of certain DNA nanostructures\c{barish2009information}, allows growth to proceed at values of the chemical potential $\mu$ too large to see direct nucleation within the simulation box. No move of any block in the template layer was permitted (i.e. we consider it to be a rigid structure whose only purpose is to promote growth). 

To quantify the fidelity of the assembly process we defined assembly `yield' as the fraction of block types found in the exact position they occupy in the ideal structure. According to this definition of yield, the `equilibrium yield' can be less than unity (the ideal structure is the lowest-energy structure; the target structure, which is the lowest {\em free-energy} structure, can potentially harbor defects). In what follows we shall display both equilibrium yield and `dynamic yield', the latter meaning yield achieved after some elapsed time of the growth process. As we shall demonstrate, in some regimes of parameter space self-assembly is initially of poor quality, and can become much better on timescales that are longer but nonetheless accessible to our simulations (and would be accessible to the corresponding experiments). In other regimes, assembly can remain poor on timescales beyond those accessible to our simulations (or to the corresponding experiment). We define one time unit as one Monte Carlo 
`sweep', namely $q$ attempted changes of state of randomly-chosen lattice sites. 

Given that all block types in the ideal structure are distinct, it is convenient to associate with each block a pixel, and arrange for those pixels to make, in the ideal structure, an arbitrarily-chosen pattern (see \f{fig3}, top). This pattern provides a guide to the eye as to the fidelity of the assembly process.
\begin{figure*}[t!]
\centering
\includegraphics[width=0.79\linewidth]{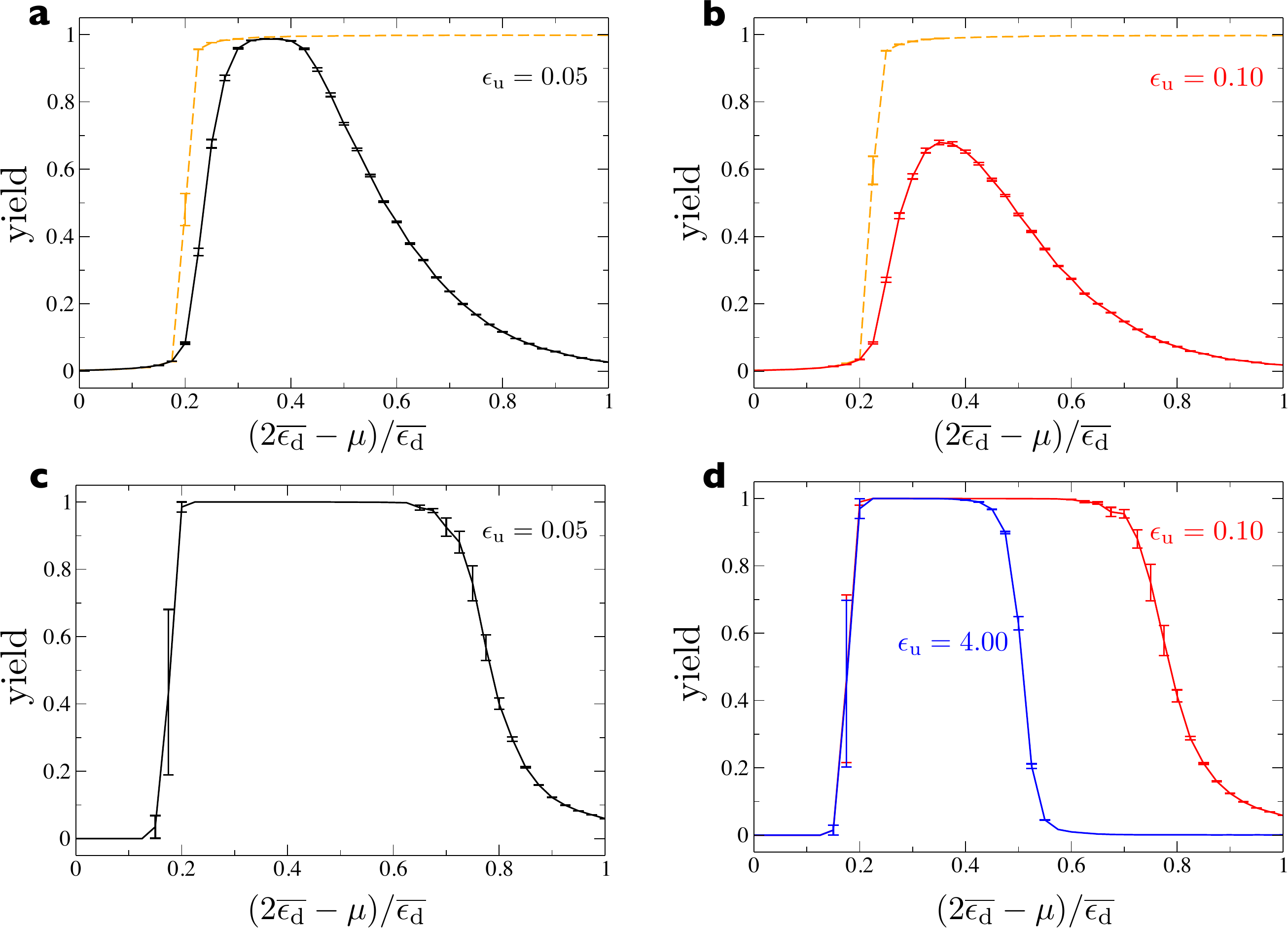}
\caption{\label{fig5} The fidelity of growth in the presence of attractive undesigned interactions of fixed strength can be improved by increasing the energy scale of the set of designed interactions (going from top to bottom). (a,b) In the top panels, undesigned (labelled) and designed interactions ($\en=4$) are not well-separated in scale (note that $\ln q \approx 7.3$), and the window of good assembly is relatively narrow. Here, growth results in a collection of randomly-arranged block types from which the target structure grows only subsequently (see \f{fig_time_series2}). Dashed lines show the stability of the equilibrium structure after fixed time. (c,d) In the bottom panels, undesigned (labelled) and designed interactions ($\en = 11.3$) are well-separated in energy scale, and the window of good assembly is wider for the two choices of $\enn$ used in the top panels. The window is also reasonably wide for a much larger undesigned attraction of $\enn=4$. In these cases, growth results directly in the target structure. Dynamic yield is calculated after $10^7$ time steps. Each curve is averaged over 200 independent trajectories and error bars show the standard deviation of each data point.}
\end{figure*}

\section{Results: designed interactions only}
\label{results1}

\noindent {\bf General expectations.} We began by suppressing completely all undesigned interactions, i.e. we set $\denn =0$ and $\benn \to - \infty$. We first chose all designed interactions to be equal in strength, i.e. we set $\den =0$, giving $\en = \ben$. We then studied self-assembly at various values of the chemical potential $\mu$. To estimate the likely interval or `window' of $\mu$ within which assembly will be successful, we were guided by previous analysis of lattice gas growth pathways\c{shneidman1999nucleation,shneidman2003lowest} to reason as follows. 

Looking at \f{fig2}, and focusing on the (smooth) bottom row of the growing structure, we want the addition of one designed bond to that smooth surface to be less likely than removal of that same bond, once made; otherwise, we would be near the `spinodal' limit, and fragments of the target structure would begin to appear everywhere in the simulation box. From the Monte Carlo proposal and acceptance rates of our dynamic protocol, described in Section~\ref{model}, the rate of appearance, per Monte Carlo sweep, of a designed bond at a smooth surface is $q^{-1}\min\left( 1, q\, {\rm e}^{\en- \mu} \right)$. The prefactor $q^{-1}$ arises because only one block type, out of the ensemble of $q$ block types, can make a designed bond in a given position. Once in place, and provided its environment does not change (which is a strong assumption that we expect to be untrue in general), this block disappears with rate $\min\left( 1, q^{-1}{\rm e}^{-\en+ \mu} \right)$. The ratio of these `on' to `off' rates is $\e^{\en- \mu}$, meaning that if $\mu > \en$ then blocks making only one designed 
bond 
are unstable with respect to their removal.

Looking again at \f{fig2}, we also want two designed bonds to be stable with respect to their removal, so that extended layer growth is possible. Considering the rates of appearance and removal of a block able to make two designed bonds (e.g the block of type $i+L$), we require $\mu < 2\en$ if blocks making two designed bonds are to be stable with respect to their removal. Thus, our expectation is that in the presence of designed bonds of fixed strength $\en$, the likely window of $\mu$ within which growth of the equilibrium target structure will be happen is
\beq
\label{window1}
\en < \mu < 2\en.
\eeq
Moving to one side of this window, as $\mu$ approaches and then exceeds $2 \en$, we would expect layer growth to become slow (followed by the target structure becoming unstable with respect to dissolution). Moving to the other side of the window, as $\mu$ approaches and then drops below $\en$, we would expect fragments of the target structure to appear throughout the simulation box, and for the growth process to result in a far-from-equilibrium collection of distinct and likely incommensurate pieces of the target. In principle this kinetically-trapped collection of fragments will turn into the equilibrium structure given sufficient time, but that would require the re-dissolution of assembled fragments, a process that we would expect to be very slow for fragments of appreciable size. Thus, \eqq{window1} is the window within which we would expect to see most rapid self-assembly of the equilibrium structure. \\

\noindent {\bf Simulation results.} In \f{fig3}(a) we show that this expectation is borne out, approximately, by our simulations. In these simulations we set $\en=4$ (with $\den=0$). For a large range of $\mu$ the equilibrium structure is the ideal structure with essentially no vacancies, i.e. the `equilibrium yield' is unity. Within the window of $\mu$ prescribed by \eqq{window1}, corresponding to the interval 0 to 1 on the horizontal axis of the figure, the equilibrium target structure self-assembles with high fidelity. The time series labeled by the green circle shows an example of high-fidelity assembly. Growth eventually becomes impossible as we move leftward on panel (a), first because of slow layer completion, and then because the ideal structure is not stable. As we move rightward on the figure, growth eventually results in a structure different to the equilibrium one, first because of vacancy incorporation in the growing structure, and eventually because of the nucleation of multiple fragments of the target structure within the simulation box. 

If instead we draw designed interactions from a Gaussian distribution of mean $\ben=4$ and nonzero variance $\den$, we find that \eqq{window1} -- with $\en$ replaced by $\ben$ -- remains a necessary condition for self-assembly of the equilibrium target structure, but is no longer a sufficient condition. As shown in \f{fig3}(b--d), the window in which the equilibrium target structure assembles narrows and eventually disappears entirely as the distribution from which designed interaction strengths are drawn broadens. The narrowing of the window happens from both sides: at the right-hand side, the strongest bonds act to impair assembly, causing fragments of the target structure (often repeated fragments) to appear throughout the simulation box (see time-series labeled by the red square). At the left-hand side of the window, layer growth is slowed by the weakest contacts, even though the thermodynamic stability of the target structure is enhanced by having a distribution of bond strengths. 

%The oscillations in panel (c) and (d) arise from the following effect. As $\mu$ is decreased, i.e. as the driving force for assembly is increased (moving left to right on the plots), yield first declines because the growing template encounters incommensurate clusters, formed via the strongest set of bonds, that have appeared elsewhere in the simulation box. For stronger driving forces the template grows more quickly, outpacing appearance of these nuclei, and so the yield increases slightly. For stronger driving still, nucleation of new cluster types happens and causes yield to degrease again. 

The closing of the `good assembly' window as designed interaction variance increases is suggested, in a qualitative sense, by the nature of the argument used to derive \eqq{window1} for the zero-variance case: there, we argued that good assembly could happen when the energy scale of attachment (set by the chemical potential $\mu$) lay between the energy of one bond and the energy of two bonds. For the finite-variance case, the distinction between the energy of one bond and two bonds becomes blurred as the interaction variance becomes sufficiently large (see figure panel insets).

In \f{fig4} we show yield as a function of time for simulations done at three of the parameter sets used in \f{fig3}. For designed interaction distribution of zero variance, comparison of the circle and triangle symbol sets shows that it takes about an order of magnitude longer to grow the target structure at the strong-driving side of the good-assembly window than in the middle of the window. For a designed interaction distribution of large variance, we see from the square parameter set that the target structure cannot be grown on accessible timescales.

We conclude from \f{fig3} and \f{fig4} that it is beneficial to draw designed interactions from as narrow a distribution as possible: in Section~\ref{discussion} we discuss the possible implications of this observation for DNA-mediated self-assembly.

\section{Results: designed and undesigned interactions}
\label{results2}
\noindent {\bf General expectations.} We next considered assembly in the presence of attractive `undesigned' interactions, and we focused on the case in which $\en$ and $\enn$ are drawn from distributions with zero variance. Looking at \f{fig2}, we estimate that a necessary condition for undesigned bonds {\em not} to become incorporated in the growing target structure (in the limit of low growth rate) is that two undesigned bonds be unstable with respect to their removal, so preventing their participation in layer growth. This rate of appearance, per Monte Carlo sweep, of a block making two undesigned bonds at a corner of the growing target structure (say, between block types $i-1$ and $i+L$) is $(1-1/q)\min\left( 1, q\, {\rm e}^{2 \enn- \mu} \right)$. Once in place, and provided that its environment does not change, this block will detach with rate $\min\left( 1, q^{-1}{\rm e}^{-2 \enn+ \mu} \right)$. The ratio of these `on' to `off' rates is $(q-1) {\rm e}^{2 \enn- \mu}$. For two designed bonds to be 
unstable to their removal we must therefore have
\beq
\label{c2}
\mu > 2 \enn + \ln(q-1).
\eeq
In a sense, one can regard the term $\ln(q-1)$ on the right-hand side of \eqq{c2} as an effective `entropic stickiness' that results from the fact that `incorrect' block types are numerous (recall that we have scaled $\mu$ and $\enn$ by $\kt$ already; if \eqq{c2} was rewritten in units with dimensions, then the term $\ln(q-1)$ would come with a prefactor $\kt$).

\begin{figure}[]
\centering
\includegraphics[width=\linewidth]{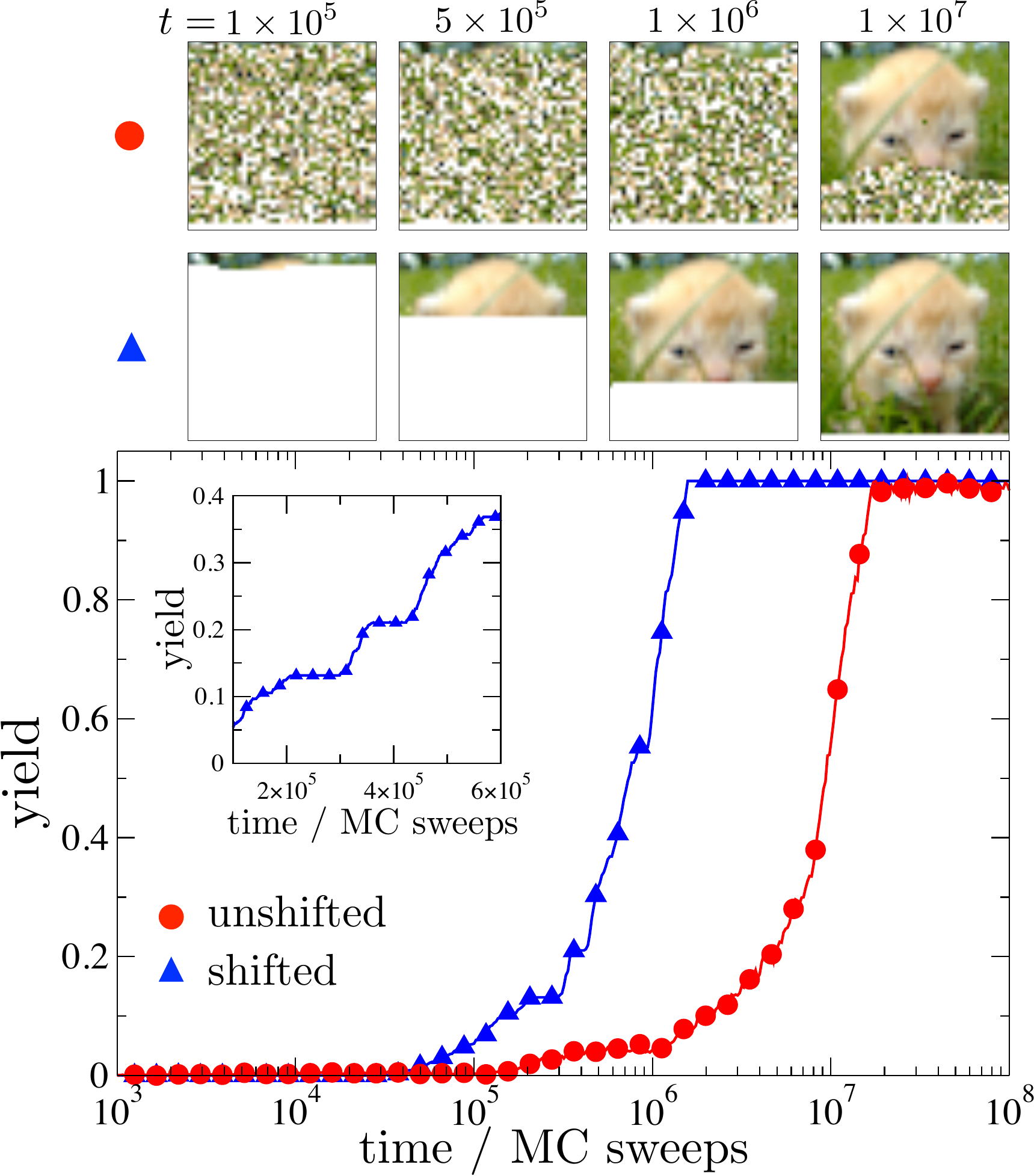}
\caption{\label{fig_time_series2} Yield-versus-time plots for examples of growth using undesigned and designed interactions that are respectively well-separated in energy (`shifted'; triangles) and poorly-separated in energy (`unshifted'; circles); see \f{fig5}. The well-separated energy scale results in the direct appearance of the equilibrium target structure. The inset shows that the target structure grows in this regime in a layer-by-layer fashion.}
\end{figure} 
\begin{figure*}[]
\centering
\includegraphics[width=\linewidth]{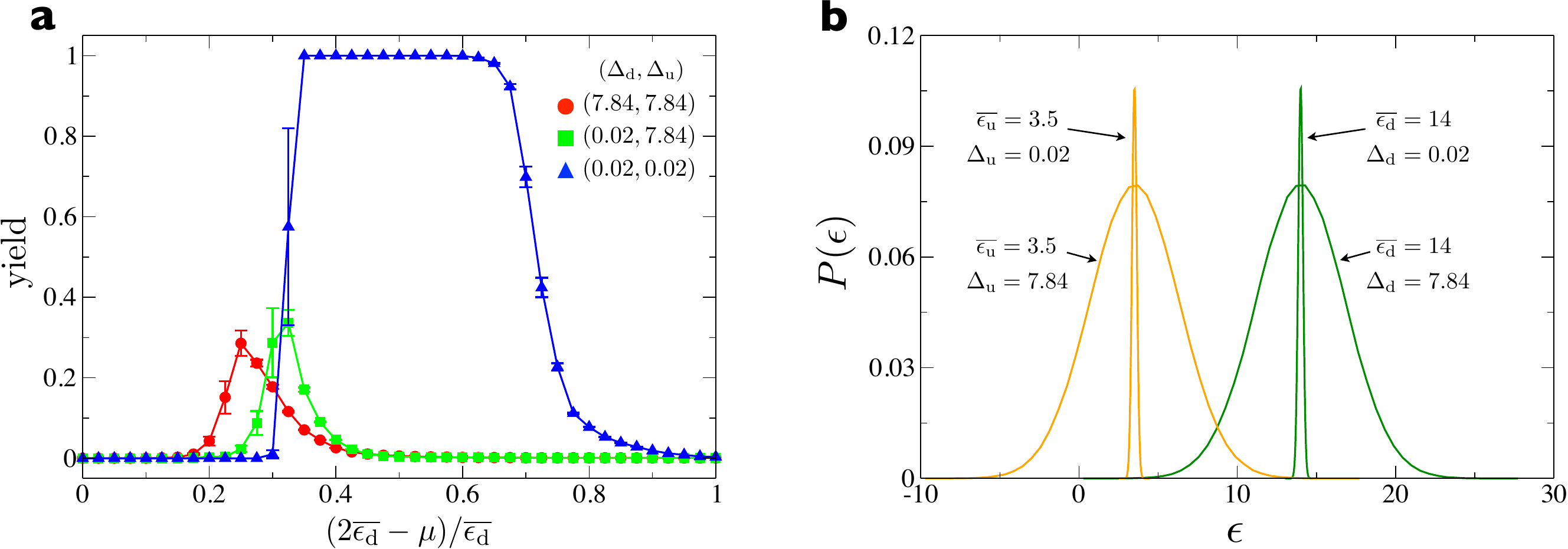}
\caption{\label{fig_both} Yield-versus-time plots for examples of growth using undesigned and designed interaction distributions of a range of widths. (a) Only when both distributions are relatively narrow and well-separated in energy (blue triangles) is assembly good within a substantial window; widening one (green squares) or both distributions (red circles) impairs yield. Each curve has been averaged over 200 independent trajectories and error bars indicate the standard deviation at each data point. (b) Distributions used to generate panel (a).}
\end{figure*} 

We therefore estimate that if \eqq{window1} and \eqq{c2} hold simultaneously, growth should result in the equilibrium target structure. It is straightforward in principle to satisfy these two relations simultaneously. Say that the number of block types $q$ and the undesigned interaction strength $\enn$ are fixed. \eqq{window1} and \eqq{c2} predict that the equilibrium structure should grow if $\en$ can be made large enough that the hierarchy 
\beq
\label{window2}
 2 \enn + \ln(q-1) < \mu <2\en
 \eeq
 can be arranged, and that the {\em largest} window of good assembly can be achieved if $\en$ is large enough that the hierarchy
\beq
\label{window3}
 2 \enn + \ln(q-1) < \en < \mu <2\en
\eeq
can be arranged. Equations \eq{window2} and \eq{window3} suggest that, even in the presence of attractive undesigned interactions, assembly of the equilibrium structure will occur if one makes designed interactions strong enough, and works at low enough block concentrations (i.e. large enough $\mu$). 

An alternative way of viewing \eqq{window2} is to recognize that it constrains, approximately, a hierarchy of bulk free energies. The (reduced) bulk free-energy density of an `undesigned' structure, by which we mean a full lattice of randomly-arranged block types, is approximately $-2 \enn - \ln q$. The first term comes about because each particle in the structure has a half share of four undesigned bonds; the second term accounts for the entropy of mixing of component types in the grand-canonical ensemble. The bulk free-energy density of the solvated phase is $-\mu$~\footnote{See note in Section~\ref{model}: changing $q$ without changing $\mu$ results in an effective increase of the solution concentration of blocks. $\mu$ must be increased by an amount $\ln q$ if one wishes to increase the number of block types at fixed notional block concentration.}. The bulk free-energy density of the designed structure is $-2 \en$. Thus, \eqq{window2} states, approximately, that the bulk free-energy density of undesigned structures should be larger than the bulk free-energy density of the solvated phase. If, instead, the free-energy density of undesigned structures lie intermediate between the free energy densities of the solvated phase and the target structure, then one might expect that undesigned structures, and not the target, would be the first thing to grow from solution. This expectation is based on the assumption that undesigned structures, if viable thermodynamically, should be more accessible kinetically than the designed one: a fraction $1-q^{-1}$ of all block-block encounters will cause undesigned bonds to appear, but only 1 in $q$ encounters will result in a designed bond. Subsequently, one might expect to see undesigned structures evolve into the target structure\c{cardew1984kinetics,threlfall2003structural,ostwald1897studies} on some time controlled by the basic timescale $\exp(4 \enn-\mu)$ for removal of blocks from the bulk of an undesigned structure. Given that this timescale can be large, it would seem to be a better strategy to arrange for the target structure to be the first thing to grow from solution.

Thus, if \eqq{window2} holds then we expect growth to result in the equilibrium structure. If, however, Equation~\ref{window2}'s first inequality is reversed then we might expect the initial result of growth to be a dense phase of randomly-arranged block types (from which the target structure may emerge on some longer timescale).\\

\noindent {\bf Simulation results.} In \f{fig5} we see that the qualitative expectations implied by Equations~(\ref{c2})--(\ref{window3}) are borne out in simulations: assembly can be improved markedly by increasing the energy scale of designed interactions in the face of attractive undesigned interactions. Panels (a) and (b) of \f{fig5} show that for two (small) values of $\enn$, namely $\enn = 0.05$, and $\enn = 0.1$, self-assembly of the target structure is less successful than it was in \f{fig3}(a), for the same designed-interaction energy scale ($\en = 4$) and the same timescale. Equation \eq{window3} suggests that the problem with these parameter choices is that the effective energy scales $2 \enn + \ln(q-1)$ and $2 \en$ are not well-enough separated: the term $\ln(q-1)$ alone is approximately 7.35. As a result, growth results in a collection of randomly-arranged block types, from which the target structure emerges only subsequently. Thus, in \f{fig5}(c) we have increased $\en$ to $11.3$. As predicted, the window of good self-assembly (which now occurs at larger $\mu$, i.e. smaller notional block concentration) widens considerably.  

The yield-versus-time plots shown in \f{fig_time_series2} demonstrate the difference in the rate of attainment of good yield for well-separated and poorly-separated undesigned and designed interaction energy scales. Growth using the smaller (`unshifted') designed-interaction energy scale results in the formation of a randomly-arranged set of blocks, from which the target structure eventually emerges. Note that the characteristic time for a collection of randomly-arranged component types to spawn the target structure must scale roughly as $\exp(4 \enn-\mu)$, the rate to break four undesigned bonds. Growth using the larger (`shifted') designed-interaction energy scale results in direct (and more rapid) appearance of the equilibrium target structure. Growth in this regime involves slow layer-by-layer nucleation (see inset), a phenomenon seen in several real systems\c{schmit2012growth,ocko1986quantized}. 

We also performed simulations in which undesigned interactions were selected from distributions with nonzero variance (\f{fig_both}). When designed and undesigned interactions are both drawn from narrow, well-separated distributions, there exists a broad window of good self-assembly (blue triangles). However, if the variance of the undesigned interaction distribution is increased substantially, then assembly is impaired (green squares), because of persistent formation of blocks making unusually strong undesigned interactions. 

The results of this section demonstrate that direct growth of equilibrium multicomponent structures can be achieved in the face of attractive undesigned interactions, provided that the designed interaction energy scale is well-separated from the undesigned interaction energy scale.

\section{Discussion: implications for DNA-mediated self-assembly}
\label{discussion}

\begin{figure*}[t!]
\centering
\includegraphics[width=0.82\linewidth]{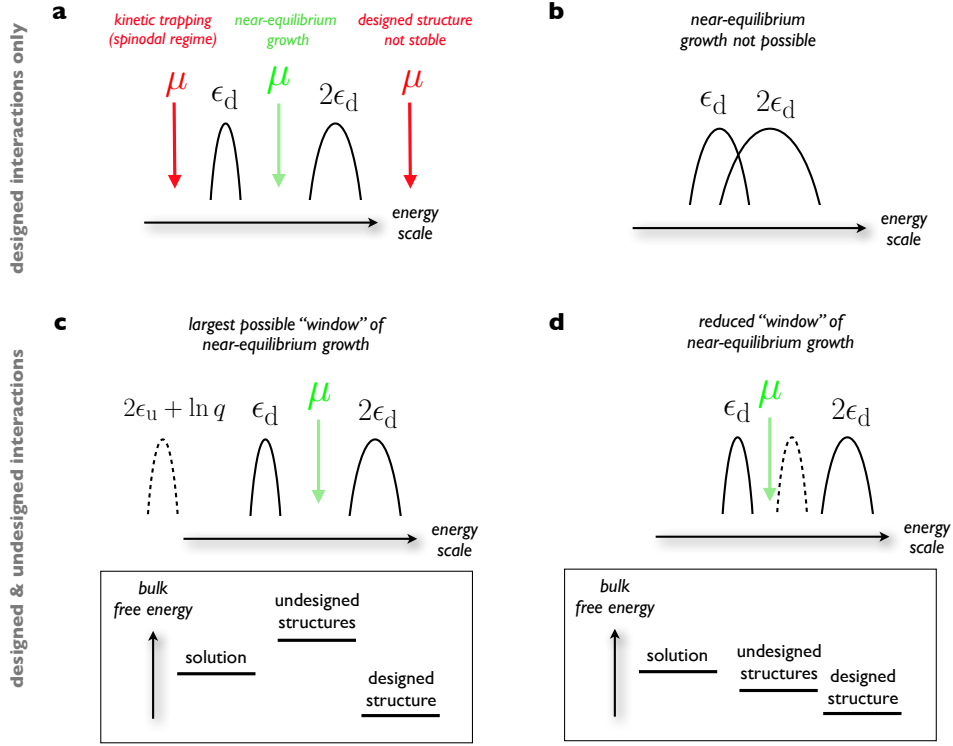}
\caption{\label{fig_design} Principles for growth of equilibrium multicomponent structures derived from the results of Sections~\ref{results1} and~\ref{results2}. Panels (a) and (b) relate to the case of designed interactions only. (a) We want the basic rate for removing a particle from solution to lie intermediate between the basic rates for the breaking of one and two designed bonds. (b) If the distribution from which designed interactions are drawn is too broad, then this energy-scale separation cannot be achieved, and growth of the equilibrium structure is not possible. Panels (c) and (d) relate to the case in which undesigned interactions are present. (c) Direct growth of the equilibrium structure is possible in the face of attractive undesigned interactions, provided that the energy scales $\mu$ and $\en$ are made large enough, i.e. provided that one works with a low solution concentration of blocks and with strong designed bonds. (d) Otherwise, some (or all) of the notional window of good assembly will result in rapid formation of a randomly-arranged collection of block types (these considerations can also be viewed as constraining a hierarchy of bulk free energies: see boxes).}
\end{figure*}
\begin{figure*}[t]
\centering
\includegraphics[width=0.9\linewidth]{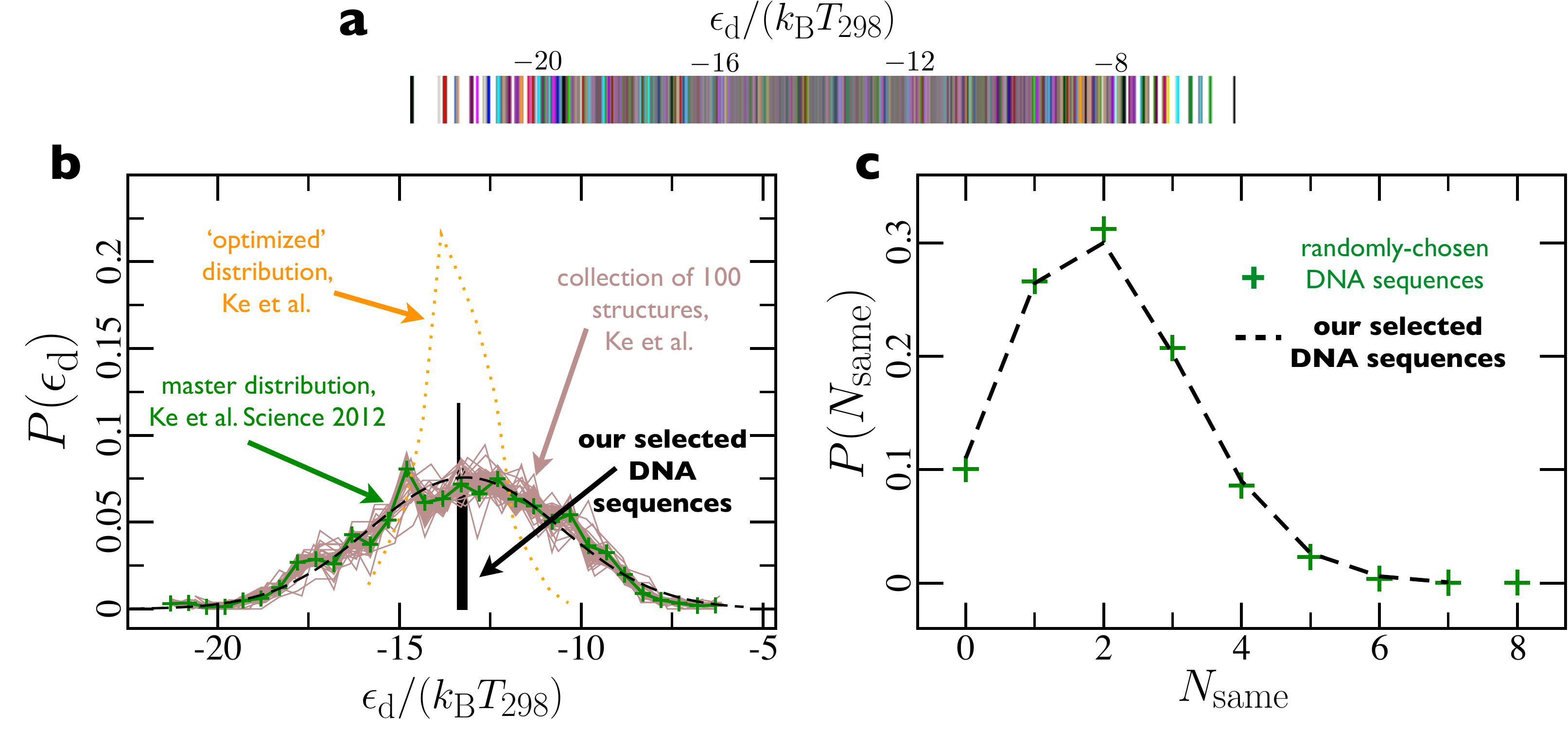}
\caption{ \label{fig_dna1} The `designed interaction' distribution of DNA `bricks'\c{ke2012three} may be substantially narrowed through rational selection of complementary DNA sequences. (a) Set of hybridization free energies for all possible non-palindromic 8-basepair complementary sequences, in units of $\kt$ at 298 K. (b) In Ref.\c{ke2012three} complementary sequences (which mediate what we call `designed interactions') are chosen randomly from this collection, leading to the green (molecular canvas) and grey (individual structure) distributions; these have a substantial variance (one set of `optimized' interactions from that work is shown in orange; this distribution is narrower than the random case, but still possesses considerable width). Individual structures in Ref.\c{ke2012three} contain an average of about $1500$ complementary sequences. If, by contrast, we choose from (a) the set of $2000$ sequences (for instance) whose interaction energies are most similar, then we can achieve a substantial narrowing of the designed interaction distribution with little change in the mean (black distribution). (c) Non-complementary sequences alike in energy are no more alike in identity than are randomly-chosen ones, suggesting that this rationally-selected set of DNA interactions would have `undesigned' interactions no more potent than those of Ref.\c{ke2012three}. Here $N_{\rm same}$ is calculated by aligning two sequences and counting the number of positions that display identical nucleotide type; this process is then done for all pairs of sequences in the set. We carried out a similar procedure to detect complementary nucleotides, and repeated both calculations for anti-aligned sequences (see supplemental files); all measures show random and purposeful selection to be similar in respect of unintended sequence complementarity.}
\end{figure*}

\noindent {\bf General principles for growth of equilibrium multicomponent structures.} The results of Sections~\ref{results1} and~\ref{results2} suggest two principles, summarized in \f{fig_design}, that must be observed in order to ensure growth of equilibrium structures built from many distinct component types. First, we need a sufficiently narrow distribution of designed interaction strengths. Second, the energy scale of designed interactions and undesigned interactions should be well-separated, so that undesigned contacts cannot proliferate at the expense of designed ones. Although these principles are suggested by the study of a two-dimensional model, we were guided to them by simple arguments relating only to the number of interactions made by building blocks as they bind to a structure. These arguments are not specific to two dimensions, and we therefore anticipate that the principles we have observed should also apply in three dimensions. Further, although we have not considered nucleation explicitly, we assume that if growth results in the equilibrium structure then nucleation will not result in something drastically different, assuming that it happens fast enough to be seen in experiment. Motivated by these assumptions, we now make suggestions for how to choose DNA sequences in order to maximize the efficiency of DNA-mediated self-assembly.\\

\noindent {\bf Selecting DNA strands for optimal `DNA brick' self-assembly.} Ke et al.\c{ke2012three} reported the self-assembly of 100 DNA structures of finite size. These structures were derived or ``sculpted'' from a ``molecular canvas'' (Tables S14 \& S15 of the Supplemental Information of Ref.\c{ke2012three}), a 3D assembly of 4,455 DNA bricks interacting via a total of 13,860 8-basepair strands. The 100 structures sculpted from this canvas were built from $416\pm85$ DNA blocks and $1506\pm326$ interacting strands (bricks in the bulk of a structure usually make 4 nearest-neighbor contacts; in what follows we ignore poly-T border strands that enforce the finite size of structures). 

In the language of the present paper, DNA bricks possess `designed interactions' mediated by complementary DNA strands 8 basepairs long. Strands were chosen to be complementary in a random fashion. We show in \f{fig_dna1}(a) the collection of interaction energies (hybridization free energies) for all possible complementary 8-basepair sequences. We computed these hybridization free energies using the model of SantaLucia et al.\c{santalucia2004thermodynamics}. In \f{fig_dna1}(b) we show that if one chooses randomly from this collection of energies then one obtains a distribution of designed interaction energies that is relatively broad (green and grey lines). These distributions are broader than those used to compute the model results in \f{fig3}(d). If, by contrast, one chooses sets of sequences {\em closest in energy}, then one can achieve a significant narrowing of this distribution (black line). This distribution is similar in width to that used to compute the model results in \f{fig3}(b). In \f{fig_dna2} we show that one can select sets of complementary DNA interactions with a similarly narrow distribution for a range of mean interaction energies, and that this can be done for a range of sequence lengths.

Importantly, non-complementary sequences alike in energy are no more alike in sequence identity than are randomly-chosen ones -- see \f{fig_dna1}(c) -- suggesting (but not proving) that choosing designed interactions in this manner would result in undesigned interactions no more potent than do randomly-chosen sequences. The results of this paper therefore suggest that this rationally-selected set of sequences will lead to better-quality self-assembly than does random sequence choice, with one important caveat:  we have no prescription for calculating accurately the energies of non-complementary sequences, and we cannot say with certainty what are the `undesigned' interactions that result from these sequences. If the latter are large, for some reason that we have not anticipated, then narrowing the distribution of designed interactions alone may not substantially improve assembly (see \f{fig_both}). Nonetheless, based on the results of our paper, these selected sequences are our best estimate for the best way of doing DNA brick self-assembly. 

In the supplemental file \href{http://nanotheory.lbl.gov/people/dna_sequences/si_sequences_n8.txt}{\texttt{si\_sequences\_n8.txt}} we list 15 sets of 2000 `narrow-distribution' non-palindromic DNA sequences of length 8 that one might use to self-assemble DNA brick structures. These sets have been chosen to have different mean interaction energies, and variances as small as possible (sets correspond to various parts of the `basin' of the green curve in \f{fig_dna2}(c)). A similar list of 15 sets of 2000 bases is provided for sequences of length 9 (\href{http://nanotheory.lbl.gov/people/dna_sequences/si_sequences_n9.txt}{\texttt{si\_sequences\_n9.txt}}). Additionally, we provide two data files whose sequences contain repeats of no more than two nucleotides (\href{http://nanotheory.lbl.gov/people/dna_sequences/si_sequences_n8_no_triples.txt}{\texttt{si\_sequences\_n8\_no\_triples.txt}}, \href{http://nanotheory.lbl.gov/people/dna_sequences/si_sequences_n9_no_triples.txt}{\texttt{si\_sequences\_n9\_no\_triples.txt}}). This restriction may help to suppress undesigned interactions, and can be arranged at the cost of only a slight increase in the variance of the distribution of designed interactions.\\

\begin{figure*}[t]
\centering
\includegraphics[width=0.9\linewidth]{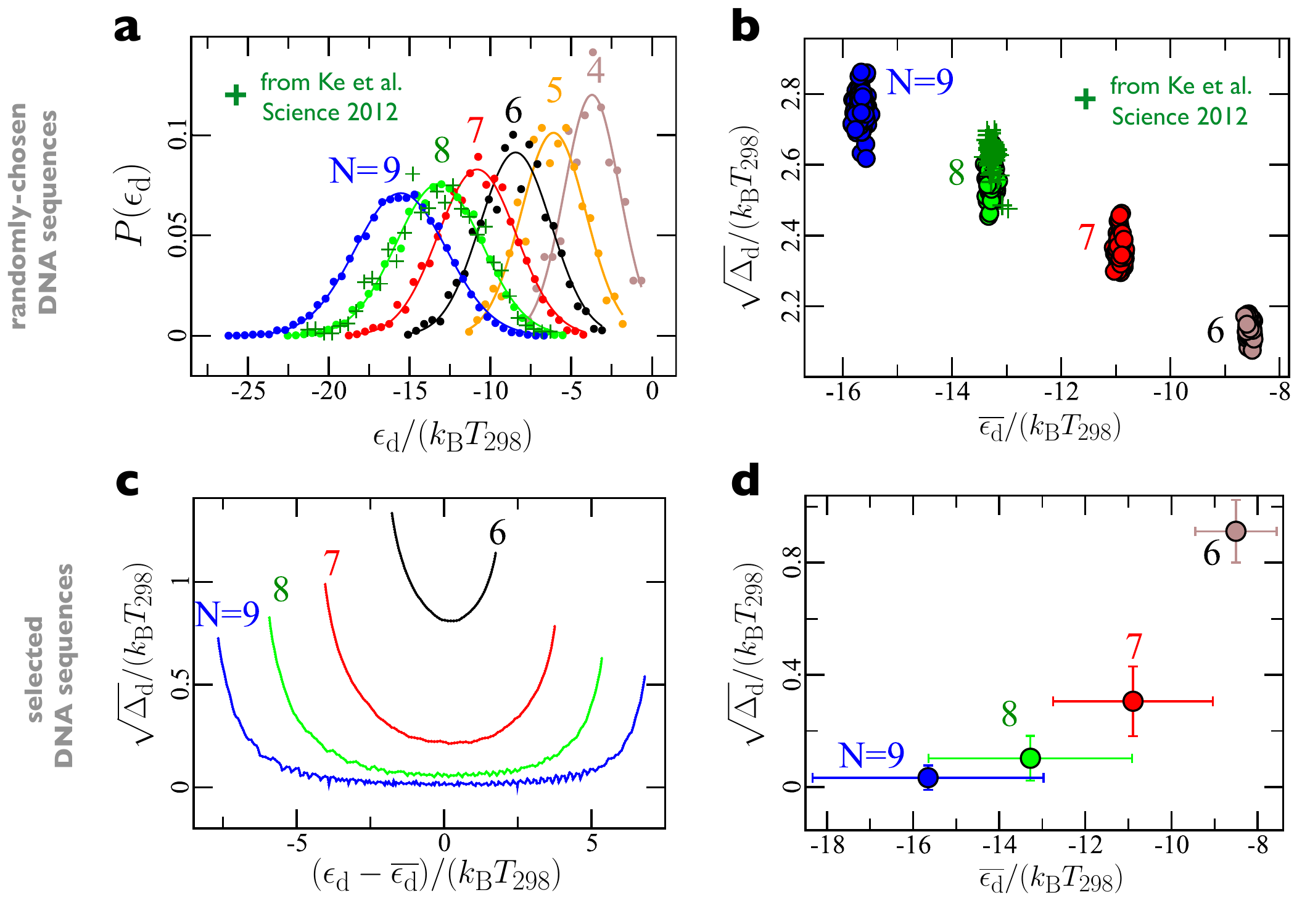}
\caption{\label{fig_dna2} Narrow-variance designed interactions can be arranged for DNA sequences of a range of lengths $N$. (a) Distributions of hybridization free energies for DNA strands of length $N$ basepairs (with their complementary strand). Solid lines are distributions of all possible non-palindromic interaction energies (hybridization free energies), while dots are distributions of 2000 sequences chosen randomly from those larger distributions. The plusses denote sequences from Ref.\c{ke2012three}. (b) Mean and variance for a number of random 2000-sequence selections from all possible sequences of length $N$: as $N$ increases, so does the interaction variance. (c) By contrast, sampling (for instance) 2000 nearest neighbors in the energy spectrum results in a much reduced variance. Here we plot the variance of 2000 sequences that are nearsest-neighbors in each energy spectrum. For $N=8$, moving left to right on the horizontal axis can be thought of as moving left to right along the `bar code' shown in \f{fig_dna1}(a), and taking the 2000 sequences closest in energy at each point. The broad basin for $N=8$ and 9 shows that many distinct small-variance 2000-basepair sets with different mean interactions can be chosen (some are listed in \href{http://nanotheory.lbl.gov/people/dna_sequences/si_sequences_n8.txt}{\texttt{si\_sequences\_n8.txt}} and \href{http://nanotheory.lbl.gov/people/dna_sequences/si_sequences_n9.txt}{\texttt{si\_sequences\_n9.txt}}. (d) Similar to (b), but for rationally-selected DNA interactions: here, the `designed interaction' variance can be arranged to {\em decrease} with increasing $N$.}
\end{figure*}
\section{Conclusions}
\label{conclusions}

We have shown within a lattice-based computer model that the high-fidelity growth of an equilibrium structure composed of a large number of precisely-arranged component types can be ensured be observing some relatively simple conditions. These conditions relate to the energies of building blocks' `designed' interactions, which stabilize energetically the unique target structure, and to the energies of their `undesigned' interactions, which allow blocks to associate in a compositionally-random way. We find that direct growth of the equilibrium structure can happen in the face of substantial attractive undesigned interactions, an observation that may explain why DNA bricks can self-assemble successfully even when attractive `undesigned' interactions are not intentionally suppressed\c{ke2012three}. We also find that best assembly happens when designed interactions are drawn from a distribution that is as narrow as possible, and we have shown that one can choose DNA sequences so as to achieve a narrow distribution of such energies. These sequences are available as supplementary files (see Section~\ref{discussion}).

Many multicomponent systems self-assemble in a kinetically trapped manner\c{kim2008probing,scarlett2011mechanistic,sanz2007evidence,peters2009competing,whitelam2012self}, particularly when the energy scales associated with component-type interactions cannot be controlled precisely. But when these interactions can be controlled, experiments\c{ke2012three} and simulations\c{halverson2013dna,reinhardt2014numerical} demonstrate that self-assembly of equilibrium structures built from $Q\sim 10^3$ distinct component types is possible. Simple scaling arguments suggest that the strategy of Ref.\c{ke2012three} will work for a range of values of $Q$ in excess of those ($Q \sim 10^3$) that have been used to date, but the assembly of truly macroscopic structures may require some modification of this basic protocol. Both thermodynamic and dynamic factors would seem to impose eventual limitations in this regard.

Thermodynamically, the requirement that the equilibrium structure is composed of an arbitrary arrangement of $Q$ distinct components requires increasingly strong interactions to achieve as $Q$ becomes larger, simply to counter the entropy gained upon mixing component types. The equilibrium density of misplaced blocks will go roughly as $Q \exp(-z \beta \Delta E)$, where $\Delta E$ is the difference in energy between `designed' and `undesigned' interactions, and $z$ is the number of bonds made by blocks. For macroscopic structures, i.e. $Q \sim 10^{24}$, and for the case $z=4$, appropriate to the experiments of \c{ke2012three}, one would need $\Delta E$ to approach about $30\, \kt$ before one has of order unity defects in the equilibrium structure. This energy scale is achievable using DNA-like interactions, but, if undesigned interactions are not repulsive, implies attractive interactions almost as strong as covalent bonds. 

Dynamically, the basic timescale for the growth of large$-Q$ structures is large. In our simulations the basic timescale for growth of the target structure scales as $Q$, because only about 1 in every $Q$ block-block encounters results in the creation of a designed bond. A similar scaling would seem likely in experiment. Assuming a basic block attachment timescale of $10^{-10}$ s, which is roughly the time taken by a small molecule to diffuse its own diameter in water, the basic binding timescale in the presence of $10^{17}$ component types would be of order a year. Therefore, in some (very large) size limit it would seem that the basic protocol studied here must be modified somehow, e.g. by using strongly repulsive undesigned interactions, or long-ranged designed interactions that bring selected blocks together from afar. 

\section{Acknowledgements}

We thank Yonggang Ke and Peng Yin for correspondence and for sending us a file containing DNA sequences used in Ref.\c{ke2012three}. This work was done at the Molecular Foundry at Lawrence Berkeley National Lab, and was supported by the Office of Science, Office of Basic Energy Sciences, of the U.S. Department of Energy under Contract No. DE-AC02--05CH11231. This research used resources of the National Energy Research Scientific Computing Center, which is supported by the Office of Science of the U.S. Department of Energy under Contract No. DE-AC02-05CH11231.\\

Image permissions, \f{fig1}: Image in column 1 of panel (a) is reprinted with permission from Ref.\c{sanz2007evidence} (hyperlink \href{http://journals.aps.org/prl/abstract/10.1103/PhysRevLett.99.055501}{here}) copyright 2007 by  The American Physical Society. Image in column 1 of panel (b) is taken from Wikipedia\c{nacl}.  Image in column 1 of panel (c) is adapted from Ref.\c{kim2008probing}, by permission of Macmillan Publishers Ltd (Nature Materials) copyright 2008. Image in column 1 of panel (d) reproduced from Ref.\c{ke2012three}, copyright 2012 by The American Association for the Advancement of Science. Image in column 1 of panel (f) reproduced from Ref.\c{kong2013mapping}, copyright 2013 by The American Association for the Advancement of Science.

%\bibliography{bib}

\begin{thebibliography}{29}
\expandafter\ifx\csname natexlab\endcsname\relax\def\natexlab#1{#1}\fi
\expandafter\ifx\csname bibnamefont\endcsname\relax
  \def\bibnamefont#1{#1}\fi
\expandafter\ifx\csname bibfnamefont\endcsname\relax
  \def\bibfnamefont#1{#1}\fi
\expandafter\ifx\csname citenamefont\endcsname\relax
  \def\citenamefont#1{#1}\fi
\expandafter\ifx\csname url\endcsname\relax
  \def\url#1{\texttt{#1}}\fi
\expandafter\ifx\csname urlprefix\endcsname\relax\def\urlprefix{URL }\fi
\providecommand{\bibinfo}[2]{#2}
\providecommand{\eprint}[2][]{\url{#2}}

\bibitem[{\citenamefont{Sanz et~al.}(2007)\citenamefont{Sanz, Valeriani,
  Frenkel, and Dijkstra}}]{sanz2007evidence}
\bibinfo{author}{\bibfnamefont{E.}~\bibnamefont{Sanz}},
  \bibinfo{author}{\bibfnamefont{C.}~\bibnamefont{Valeriani}},
  \bibinfo{author}{\bibfnamefont{D.}~\bibnamefont{Frenkel}}, \bibnamefont{and}
  \bibinfo{author}{\bibfnamefont{M.}~\bibnamefont{Dijkstra}},
  \bibinfo{journal}{Physical Review Letters} \textbf{\bibinfo{volume}{99}},
  \bibinfo{pages}{55501} (\bibinfo{year}{2007}).

\bibitem[{\citenamefont{Kim et~al.}(2008)\citenamefont{Kim, Scarlett,
  Biancaniello, Sinno, and Crocker}}]{kim2008probing}
\bibinfo{author}{\bibfnamefont{A.}~\bibnamefont{Kim}},
  \bibinfo{author}{\bibfnamefont{R.}~\bibnamefont{Scarlett}},
  \bibinfo{author}{\bibfnamefont{P.}~\bibnamefont{Biancaniello}},
  \bibinfo{author}{\bibfnamefont{T.}~\bibnamefont{Sinno}}, \bibnamefont{and}
  \bibinfo{author}{\bibfnamefont{J.}~\bibnamefont{Crocker}},
  \bibinfo{journal}{Nature Materials} \textbf{\bibinfo{volume}{8}},
  \bibinfo{pages}{52} (\bibinfo{year}{2008}).

\bibitem[{\citenamefont{Whitelam et~al.}(2014)\citenamefont{Whitelam, Hedges,
  and Schmit}}]{whitelam2014self}
\bibinfo{author}{\bibfnamefont{S.}~\bibnamefont{Whitelam}},
  \bibinfo{author}{\bibfnamefont{L.~O.} \bibnamefont{Hedges}},
  \bibnamefont{and} \bibinfo{author}{\bibfnamefont{J.~D.}
  \bibnamefont{Schmit}}, \bibinfo{journal}{Phys. Rev. Lett.}
  \textbf{\bibinfo{volume}{112}}, \bibinfo{pages}{155504}
  (\bibinfo{year}{2014}).

\bibitem[{\citenamefont{Kong et~al.}(2013)\citenamefont{Kong, Deng, Yan, Kim,
  Swisher, Smit, Yaghi, and Reimer}}]{kong2013mapping}
\bibinfo{author}{\bibfnamefont{X.}~\bibnamefont{Kong}},
  \bibinfo{author}{\bibfnamefont{H.}~\bibnamefont{Deng}},
  \bibinfo{author}{\bibfnamefont{F.}~\bibnamefont{Yan}},
  \bibinfo{author}{\bibfnamefont{J.}~\bibnamefont{Kim}},
  \bibinfo{author}{\bibfnamefont{J.~A.} \bibnamefont{Swisher}},
  \bibinfo{author}{\bibfnamefont{B.}~\bibnamefont{Smit}},
  \bibinfo{author}{\bibfnamefont{O.~M.} \bibnamefont{Yaghi}}, \bibnamefont{and}
  \bibinfo{author}{\bibfnamefont{J.~A.} \bibnamefont{Reimer}},
  \bibinfo{journal}{Science} \textbf{\bibinfo{volume}{341}},
  \bibinfo{pages}{882} (\bibinfo{year}{2013}).

\bibitem[{nac()}]{nacl}
\urlprefix\url{http://en.wikipedia.org/wiki/File:NaCl.png}.

\bibitem[{\citenamefont{Valeriani et~al.}(2005)\citenamefont{Valeriani, Sanz,
  and Frenkel}}]{valeriani2005rate}
\bibinfo{author}{\bibfnamefont{C.}~\bibnamefont{Valeriani}},
  \bibinfo{author}{\bibfnamefont{E.}~\bibnamefont{Sanz}}, \bibnamefont{and}
  \bibinfo{author}{\bibfnamefont{D.}~\bibnamefont{Frenkel}},
  \bibinfo{journal}{The Journal of chemical physics}
  \textbf{\bibinfo{volume}{122}}, \bibinfo{pages}{194501}
  (\bibinfo{year}{2005}).

\bibitem[{\citenamefont{Ke et~al.}(2012)\citenamefont{Ke, Ong, Shih, and
  Yin}}]{ke2012three}
\bibinfo{author}{\bibfnamefont{Y.}~\bibnamefont{Ke}},
  \bibinfo{author}{\bibfnamefont{L.~L.} \bibnamefont{Ong}},
  \bibinfo{author}{\bibfnamefont{W.~M.} \bibnamefont{Shih}}, \bibnamefont{and}
  \bibinfo{author}{\bibfnamefont{P.}~\bibnamefont{Yin}},
  \bibinfo{journal}{Science} \textbf{\bibinfo{volume}{338}},
  \bibinfo{pages}{1177} (\bibinfo{year}{2012}).

\bibitem[{\citenamefont{Kremer}(1978)}]{kremer1978multi}
\bibinfo{author}{\bibfnamefont{K.}~\bibnamefont{Kremer}},
  \bibinfo{journal}{Journal of Aerosol Science} \textbf{\bibinfo{volume}{9}},
  \bibinfo{pages}{243} (\bibinfo{year}{1978}).

\bibitem[{\citenamefont{Stauffer}(1976)}]{stauffer1976kinetic}
\bibinfo{author}{\bibfnamefont{D.}~\bibnamefont{Stauffer}},
  \bibinfo{journal}{Journal of Aerosol Science} \textbf{\bibinfo{volume}{7}},
  \bibinfo{pages}{319} (\bibinfo{year}{1976}).

\bibitem[{\citenamefont{Trinkaus}(1983)}]{PhysRevB.27.7372}
\bibinfo{author}{\bibfnamefont{H.}~\bibnamefont{Trinkaus}},
  \bibinfo{journal}{Phys. Rev. B} \textbf{\bibinfo{volume}{27}},
  \bibinfo{pages}{7372} (\bibinfo{year}{1983}).

\bibitem[{\citenamefont{Schmelzer et~al.}(2004)\citenamefont{Schmelzer, Abyzov,
  and M{\"o}ller}}]{schmelzer2004nucleation}
\bibinfo{author}{\bibfnamefont{J.}~\bibnamefont{Schmelzer}},
  \bibinfo{author}{\bibfnamefont{A.}~\bibnamefont{Abyzov}}, \bibnamefont{and}
  \bibinfo{author}{\bibfnamefont{J.}~\bibnamefont{M{\"o}ller}},
  \bibinfo{journal}{The Journal of Chemical Physics}
  \textbf{\bibinfo{volume}{121}}, \bibinfo{pages}{6900} (\bibinfo{year}{2004}).

\bibitem[{\citenamefont{Schmelzer et~al.}(2000)\citenamefont{Schmelzer,
  Schmelzer~Jr, and Gutzow}}]{schmelzer2000reconciling}
\bibinfo{author}{\bibfnamefont{J.}~\bibnamefont{Schmelzer}},
  \bibinfo{author}{\bibfnamefont{J.}~\bibnamefont{Schmelzer~Jr}},
  \bibnamefont{and} \bibinfo{author}{\bibfnamefont{I.}~\bibnamefont{Gutzow}},
  \bibinfo{journal}{The Journal of Chemical Physics}
  \textbf{\bibinfo{volume}{112}}, \bibinfo{pages}{3820} (\bibinfo{year}{2000}).

\bibitem[{\citenamefont{Scarlett et~al.}(2010)\citenamefont{Scarlett, Crocker,
  and Sinno}}]{scarlett2010computational}
\bibinfo{author}{\bibfnamefont{R.}~\bibnamefont{Scarlett}},
  \bibinfo{author}{\bibfnamefont{J.}~\bibnamefont{Crocker}}, \bibnamefont{and}
  \bibinfo{author}{\bibfnamefont{T.}~\bibnamefont{Sinno}},
  \bibinfo{journal}{The Journal of Chemical Physics}
  \textbf{\bibinfo{volume}{132}}, \bibinfo{pages}{234705}
  (\bibinfo{year}{2010}).

\bibitem[{\citenamefont{Scarlett et~al.}(2011)\citenamefont{Scarlett, Ung,
  Crocker, and Sinno}}]{scarlett2011mechanistic}
\bibinfo{author}{\bibfnamefont{R.}~\bibnamefont{Scarlett}},
  \bibinfo{author}{\bibfnamefont{M.}~\bibnamefont{Ung}},
  \bibinfo{author}{\bibfnamefont{J.}~\bibnamefont{Crocker}}, \bibnamefont{and}
  \bibinfo{author}{\bibfnamefont{T.}~\bibnamefont{Sinno}},
  \bibinfo{journal}{Soft Matter} \textbf{\bibinfo{volume}{7}},
  \bibinfo{pages}{1912} (\bibinfo{year}{2011}).

\bibitem[{\citenamefont{Peters}(2009)}]{peters2009competing}
\bibinfo{author}{\bibfnamefont{B.}~\bibnamefont{Peters}}, \bibinfo{journal}{The
  Journal of Chemical Physics} \textbf{\bibinfo{volume}{131}},
  \bibinfo{pages}{244103} (\bibinfo{year}{2009}).

\bibitem[{\citenamefont{Whitelam et~al.}(2012)\citenamefont{Whitelam, Schulman,
  and Hedges}}]{whitelam2012self}
\bibinfo{author}{\bibfnamefont{S.}~\bibnamefont{Whitelam}},
  \bibinfo{author}{\bibfnamefont{R.}~\bibnamefont{Schulman}}, \bibnamefont{and}
  \bibinfo{author}{\bibfnamefont{L.}~\bibnamefont{Hedges}},
  \bibinfo{journal}{Physical Review Letters} \textbf{\bibinfo{volume}{109}},
  \bibinfo{pages}{265506} (\bibinfo{year}{2012}).

\bibitem[{\citenamefont{Reinhardt and Frenkel}(2014)}]{reinhardt2014numerical}
\bibinfo{author}{\bibfnamefont{A.}~\bibnamefont{Reinhardt}} \bibnamefont{and}
  \bibinfo{author}{\bibfnamefont{D.}~\bibnamefont{Frenkel}},
  \bibinfo{journal}{arXiv preprint arXiv:1402.6228}  (\bibinfo{year}{2014}).

\bibitem[{\citenamefont{Halverson and Tkachenko}(2013)}]{halverson2013dna}
\bibinfo{author}{\bibfnamefont{J.~D.} \bibnamefont{Halverson}}
  \bibnamefont{and} \bibinfo{author}{\bibfnamefont{A.~V.}
  \bibnamefont{Tkachenko}}, \bibinfo{journal}{Physical Review E}
  \textbf{\bibinfo{volume}{87}}, \bibinfo{pages}{062310}
  (\bibinfo{year}{2013}).

\bibitem[{\citenamefont{Licata and Tkachenko}(2006)}]{licata2006errorproof}
\bibinfo{author}{\bibfnamefont{N.~A.} \bibnamefont{Licata}} \bibnamefont{and}
  \bibinfo{author}{\bibfnamefont{A.~V.} \bibnamefont{Tkachenko}},
  \bibinfo{journal}{Physical Review E} \textbf{\bibinfo{volume}{74}},
  \bibinfo{pages}{041406} (\bibinfo{year}{2006}).

\bibitem[{kit()}]{kitty}
\urlprefix\url{http://commons.wikimedia.org/wiki/File:Youngkitten.JPG}.

\bibitem[{\citenamefont{Barish et~al.}(2009)\citenamefont{Barish, Schulman,
  Rothemund, and Winfree}}]{barish2009information}
\bibinfo{author}{\bibfnamefont{R.~D.} \bibnamefont{Barish}},
  \bibinfo{author}{\bibfnamefont{R.}~\bibnamefont{Schulman}},
  \bibinfo{author}{\bibfnamefont{P.~W.} \bibnamefont{Rothemund}},
  \bibnamefont{and} \bibinfo{author}{\bibfnamefont{E.}~\bibnamefont{Winfree}},
  \bibinfo{journal}{Proceedings of the National Academy of Sciences}
  \textbf{\bibinfo{volume}{106}}, \bibinfo{pages}{6054} (\bibinfo{year}{2009}).

\bibitem[{\citenamefont{Shneidman et~al.}(1999)\citenamefont{Shneidman,
  Jackson, and Beatty}}]{shneidman1999nucleation}
\bibinfo{author}{\bibfnamefont{V.}~\bibnamefont{Shneidman}},
  \bibinfo{author}{\bibfnamefont{K.}~\bibnamefont{Jackson}}, \bibnamefont{and}
  \bibinfo{author}{\bibfnamefont{K.}~\bibnamefont{Beatty}},
  \bibinfo{journal}{Physical Review B} \textbf{\bibinfo{volume}{59}},
  \bibinfo{pages}{3579} (\bibinfo{year}{1999}).

\bibitem[{\citenamefont{Shneidman}(2003)}]{shneidman2003lowest}
\bibinfo{author}{\bibfnamefont{V.~A.} \bibnamefont{Shneidman}},
  \bibinfo{journal}{Journal of Statistical Physics}
  \textbf{\bibinfo{volume}{112}}, \bibinfo{pages}{293} (\bibinfo{year}{2003}).

\bibitem[{\citenamefont{Cardew et~al.}(1984)\citenamefont{Cardew, Davey, and
  Ruddick}}]{cardew1984kinetics}
\bibinfo{author}{\bibfnamefont{P.}~\bibnamefont{Cardew}},
  \bibinfo{author}{\bibfnamefont{R.}~\bibnamefont{Davey}}, \bibnamefont{and}
  \bibinfo{author}{\bibfnamefont{A.}~\bibnamefont{Ruddick}},
  \bibinfo{journal}{Journal of the Chemical Society, Faraday Transactions 2}
  \textbf{\bibinfo{volume}{80}}, \bibinfo{pages}{659} (\bibinfo{year}{1984}).

\bibitem[{\citenamefont{Threlfall}(2003)}]{threlfall2003structural}
\bibinfo{author}{\bibfnamefont{T.}~\bibnamefont{Threlfall}},
  \bibinfo{journal}{Organic Process Research \& Development}
  \textbf{\bibinfo{volume}{7}}, \bibinfo{pages}{1017} (\bibinfo{year}{2003}),
  ISSN \bibinfo{issn}{1083-6160}.

\bibitem[{\citenamefont{Ostwald}(1897)}]{ostwald1897studies}
\bibinfo{author}{\bibfnamefont{W.}~\bibnamefont{Ostwald}}, \bibinfo{journal}{Z.
  Phys. Chem.} \textbf{\bibinfo{volume}{22}} (\bibinfo{year}{1897}).

\bibitem[{\citenamefont{Schmit and Dill}(2012)}]{schmit2012growth}
\bibinfo{author}{\bibfnamefont{J.~D.} \bibnamefont{Schmit}} \bibnamefont{and}
  \bibinfo{author}{\bibfnamefont{K.}~\bibnamefont{Dill}},
  \bibinfo{journal}{Journal of the American Chemical Society}
  \textbf{\bibinfo{volume}{134}}, \bibinfo{pages}{3934} (\bibinfo{year}{2012}).

\bibitem[{\citenamefont{Ocko et~al.}(1986)\citenamefont{Ocko, Braslau, Pershan,
  Als-Nielsen, and Deutsch}}]{ocko1986quantized}
\bibinfo{author}{\bibfnamefont{B.}~\bibnamefont{Ocko}},
  \bibinfo{author}{\bibfnamefont{A.}~\bibnamefont{Braslau}},
  \bibinfo{author}{\bibfnamefont{P.~S.} \bibnamefont{Pershan}},
  \bibinfo{author}{\bibfnamefont{J.}~\bibnamefont{Als-Nielsen}},
  \bibnamefont{and} \bibinfo{author}{\bibfnamefont{M.}~\bibnamefont{Deutsch}},
  \bibinfo{journal}{Physical review letters} \textbf{\bibinfo{volume}{57}},
  \bibinfo{pages}{94} (\bibinfo{year}{1986}).

\bibitem[{\citenamefont{SantaLucia~Jr and
  Hicks}(2004)}]{santalucia2004thermodynamics}
\bibinfo{author}{\bibfnamefont{J.}~\bibnamefont{SantaLucia~Jr}}
  \bibnamefont{and} \bibinfo{author}{\bibfnamefont{D.}~\bibnamefont{Hicks}},
  \bibinfo{journal}{Annu. Rev. Biophys. Biomol. Struct.}
  \textbf{\bibinfo{volume}{33}}, \bibinfo{pages}{415} (\bibinfo{year}{2004}).

\end{thebibliography}

\end{document}